\documentclass[showpacs,superscriptaddress,floatfix,eqsecnum,aps,nofootinbib]{revtex4}
\usepackage{epsfig}

\usepackage{graphicx}
\usepackage{amsmath, amssymb}

\newcommand{\md}{\mathrm{d}}
\newcommand{\me}{\mathrm{e}}
\newcommand{\mi}{\mathrm{i}}
\newcommand{\mca}{\mathcal{A}}
\newcommand{\mcb}{\mathcal{B}}
\newcommand{\mcf}{\mathcal{F}}
\newcommand{\mcl}{\mathcal{L}}
\newcommand{\mcz}{\mathcal{Z}}

\DeclareMathOperator{\imag}{Im}

\begin{document}

\title{The clash of symmetries in a Randall-Sundrum-like spacetime.}
\author{Gareth Dando}\email{g.dando@unimelb.edu.au}
\affiliation{School of Physics, Research Centre for High Energy
Physics, The University of Melbourne, Victoria 3010, Australia}
\author{Aharon Davidson}\email{davidson@bgumail.bgu.ac.il}
\affiliation{Physics Department, Ben-Gurion University of the Negev, Beer-Sheva 84105, Israel}
\author{Damien P. George}\email{d.george@physics.unimelb.edu.au}
\affiliation{School of Physics, Research Centre for High Energy
Physics, The University of Melbourne, Victoria 3010, Australia}
\author{Raymond R. Volkas}\email{r.volkas@physics.unimelb.edu.au}
\affiliation{School of Physics, Research Centre for High Energy
Physics, The University of Melbourne, Victoria 3010, Australia}
\author{K. C. Wali}\email{wali@phy.syr.edu}
\affiliation{Physics Department, Syracuse University, Syracuse, NY 13244-1130, USA}

\begin{abstract}
We present a toy model that exhibits
clash-of-symmetries style Higgs field kink configurations
in a Randall-Sundrum-like spacetime. 
The model has two complex scalar fields $\Phi_{1,2}$, 
with a sextic potential obeying
global U(1)$\otimes$U(1) and discrete $\Phi_1 \leftrightarrow \Phi_2$ interchange symmetries. 
The scalar fields are coupled to $4+1$ dimensional gravity endowed with a bulk
cosmological constant. We show that the coupled Einstein-Higgs field equations
have an interesting analytic solution provided the sextic potential adopts a
particular form. The $4+1$ metric is shown to be that of a smoothed-out 
Randall-Sundrum type of spacetime. The thin-brane Randall-Sundrum limit, whereby the
Higgs field kinks become step functions, is carefully defined in terms of the
fundamental parameters in the action. The ``clash of symmetries'' feature, defined
in previous papers, is manifested here through the fact that both of the U(1) symmetries
are spontaneously broken at all non-asymptotic points in the extra dimension $w$.
One of the U(1)'s is asymptotically restored as $w \to -\infty$, with the other U(1)
restored as $w \to +\infty$. The spontaneously broken discrete symmetry ensures
topological stability.  In the gauged version of this model we find new flat-space
solutions, but in the warped metric case we have been unable to find any solutions
with nonzero gauge fields.
\end{abstract}


\maketitle

\section{Introduction}

The breaking of symmetries is one of the most important issues in particle physics.
Nature has taught us the relevance of symmetry principles, especially through
Noether's connection
of global symmetries with conservation laws and the successful description of
fundamental interactions via local gauge invariance. But to reconcile symmetric
field equations with the real world, one needs that some of the
solutions to those equations have a reduced symmetry compared to the equations
themselves. This is, of course, what we mean by `spontaneous symmetry breaking (SSB)'
or `hidden symmetry'. 

The precise mechanism by which electroweak SSB happens has yet to be experimentally
determined. In the standard model (SM), the default option is to use an elementary
scalar or Higgs field doublet with a quartic potential that induces a nonzero vacuum
expectation value for one of the components. This will soon be tested at the
Large Hadron Collider through its physical-Higgs boson search programme.

The purpose of this paper is to further develop a different method of symmetry
breaking called the `clash of symmetries' \cite{clash,so10clash,gaugeclash,triniclash}
(see also \cite{pv}). This mechanism is
relevant for brane-world extensions to the regular $3+1$ dimensional standard
model \cite{rs, otherbrane}. The important development we report here is the rigorous incorporation
of gravity into one of the toy Higgs models explored in the previous
papers on this topic \cite{gaugeclash}. Since gravity is fundamental to brane-world models, this
is a very welcome step forward.

The toy model we shall analyse consists of two complex scalar fields $\Phi_{1,2}$
coupled to $4+1$ dimensional gravity endowed with a bulk cosmological constant.
The action is constructed to be invariant under global U(1)$\otimes$U(1) transformations
and the discrete $\Phi_1 \leftrightarrow \Phi_2$ interchange operation.
The Higgs potential we construct is sextic, and its specific form is driven
by our desire to find an analytic solution to the coupled Einstein-Higgs field equations.
The spacetime metric tensor we posit has the Randall-Sundrum non-factorisable
form \cite{rs}, but with the warp factor a completely smooth function of the extra
dimension coordinate. The precise Randall-Sundrum metric emerges by taking
a certain limit. The Higgs field configurations we seek are kink-like, 
associated with the spontaneously broken discrete symmetry.

We also study the gauged version of this model, extending the analysis of Ref.\cite{gaugeclash}.
We find additional non-trivial flat-space solutions, but despite an
extensive numerical search we have been unable to find any warped metric
solutions with nonzero gauge fields.

Section \ref{clashreview} reviews the clash of symmetries mechanism, while Sec.\ 
\ref{model_ungauged} defines and analyses the ungauged version of our toy model. 
This is followed by an account of the gauged cases in Sec.\ \ref{model_gauged}.
The final section is a conclusion and
outlook for future work.

\section{The clash of symmetries}
\label{clashreview}

The default SSB mechanism employed in the SM and many extensions to it utilises
a simple spatially homogeneous, static and stable solution to the field
equations: the Higgs fields $\phi_i(x^{\mu})$ are set equal to those constant
values corresponding to an absolute minimum of the Higgs potential. The
locations of the nonzero constants within the Higgs multiplets then determine
the stability group of the vacuum state, what we commonly call the unbroken
symmetry after SSB. The nonzero Higgs fields become constant background scalar
fields throughout the universe.

There are other types of solutions to Higgs field equations that can also
serve as stable, static background fields: topological solitons, such as
kinks, strings and monopoles. While there is much interest in such
configurations, they are not usually considered as viable candidates for
playing a major role in spontaneous symmetry breaking. This is simply
because they are spatially inhomogeneous. The energy densities
required for electroweak and higher symmetry breaking are so high as to
be incompatible with the strong evidence for a universe that is spatially homogeneous
at large scales. This objection, however, does not apply to brane-world
models, because the non-trivial spatial dependence can be restricted to
the extra dimension coordinates only.

For definiteness, consider the case of one extra dimension described
by Cartesian coordinate $w$ and topologically-stable Higgs configurations
$\phi_i(w)$, some of which have kink-form with respect to $w$. The pattern of spontaneous breaking
then effectively becomes a function of $w$. The $3+1$-brane is located
at, say, $w=0$. If the degrees of freedom that are confined to the 
brane \cite{local}
are confined absolutely, then the unbroken symmetry is the stability
group of $\phi_i(w=0)$, to be denoted by $H(w=0)$ from now on (and let the
internal symmetry group of the action be $G$). 
The symmetry breaking will be communicated to
the brane-world denizens through appropriate interactions. But in a
quantal (and perhaps even in a classical) world, one would not expect 
strict confinement to the brane:
there should be some leakage off it into the bulk. One therefore expects
the brane-localised fields to also couple, but with reduced strength, to
$\phi_i(w)$ states where $w$ is slightly different from zero. If the 
stability group of $\phi_i(0<|w|<\epsilon)$, call it $H(w\sim\epsilon)$, 
is different from $H(w=0)$,
then a rather rich effective symmetry breaking outcome could ensue for
the brane world. For example, a hierarchical breaking pattern 
$G \to H(w=0) \to H(w\sim\epsilon)$ was speculated upon in Ref.\cite{so10clash},
where in that case $G=\ $SO(10), $H(w=0)=\ $SU(4)$\otimes$SU(2)$\otimes$U(1)
and $H(w\sim\epsilon)=\ $SU(3)$\otimes$SU(2)$\otimes$U(1)$^2$. The potential
application to grand unified models and the gauge hierarchy puzzle is clear.

The `clash of symmetries' phenomenon allows the unbroken symmetry at 
non-asymptotic points in the extra dimension, $|w|<\infty$, to be smaller
than the symmetry holding at $|w|=\infty$ \cite{clash,so10clash,gaugeclash,triniclash}. 
The boundary conditions
at infinity for kink configurations are set to be the Higgs vacuum
expectation values (VEVs). They force the spontaneous breaking $G \to H(|w|\to \infty)$.
In usual SSB, the Higgs VEV configuration is used throughout the whole
of space, so the symmetry is broken to the group we are calling 
$H(|w|\to \infty)$ everywhere,
homogeneously. In the clash of symmetries mechanism, this symmetry breaking
pattern only holds asymptotically with respect to the extra dimension(s).

The clash of symmetries can occur if the isomorphic subgroups left unbroken
at $w = -\infty$ and $w = + \infty$, $H(w\to -\infty) \equiv H(-\infty)$
and $H(w\to +\infty) \equiv H(+\infty)$ respectively, can be {\it differently
embedded} within the parent group $G$. In the cases examined so far in the
literature, the breakdown at finite values of $w$ is to the {\it intersection}
of the asymptotic stability groups,
\begin{equation}
H(|w|<\infty) = H(-\infty) \cap H(+\infty) \equiv H_{\rm clash}.
\end{equation}
Although $H(-\infty)$ and $H(+\infty)$ are isomorphic, their different
embeddings within $G$ leads their intersection to be a smaller 
group.\footnote{There will also typically be kink configurations that
interpolate between {\it identically} embedded subgroups. Such
kinks do not display the clash of symmetries, and one must
ensure that they have higher energy than those displaying the clash.}
Some special values of $w$, most often $w=0$, may correspond to 
stability groups larger than $H_{\rm clash}$ because the spatially
varying Higgs configurations may instantaneously pass through special
patterns as $w$ varies. For instance, a component of a Higgs multiplet
may be an odd function of $w$ such as $\tanh(w)$, so it would vanish
at $w=0$ and thus the unbroken symmetry would be enhanced at that point.
This possibility lies behind our remarks two paragraphs above about
an application to the gauge hierarchy problem.

Discrete symmetries also play a vital role: their spontaneous breakdown
leads the vacuum manifold to have disconnected pieces. Kinks interpolate
between vacuum states from disconnected pieces. Within a given topological
class of kinks, the one with lowest energy is then guaranteed to be
(topologically) stable.

The easiest example of the clash of symmetries uses $G =\ $SU(3) spontaneously
breaking to SU(2). An SU(2) subgroup can be embedded in three different
ways in SU(3): $I$-spin, $U$-spin and $V$-spin. Clash of symmetries kinks
interpolate between VEVs respecting $(I,U)$, $(U,V)$ or $(V,I)$ at 
$w=(-\infty,+\infty)$. The differently embedded asymptotic SU(2) symmetries
``clash'' at non-asymptotic points, leading $H_{\rm clash}$ to be
SU(2)$_I \cap$SU(2)$_U = \{1\}$ and so on.\footnote{To actually implement
this simple SU(3) scheme requires more work: see Ref.\cite{clash} for details.}

The model-building promise of the clash of symmetries lies in the greater
symmetry breaking power of spatially inhomogeneous Higgs configurations over
the simple homogeneous alternative of standard SSB. We hope that this will
lead to models with simpler Higgs sectors and fewer associated 
parameters.
Some of us have already speculated that the ultimate application might be
to an $E_6$ model, because of the fact that there are three different
$E_6$ generators that can serve as electric charge, corresponding to three
different embeddings of electromagnetism within the full group \cite{clash}. Standard
$E_6$ models can only use one at a time. Even more speculatively, a 
connection between that threefold structure and the three families of
quarks and leptons might 
exist\footnote{There is some similarity between the clash of symmetries mechanism
and orbifold symmetry breaking \cite{orbifold}.  In particular, Ref.\ \cite{buchmuller} 
constructs a model
where different subgroups of SO(10) are left unbroken on different branes, with
the overall unrboken symmetry being the intersection of those individual
subgroups, and with families also distributed amongst the branes.  While
there are similarities, it is clear that the mechanisms are not identical, because
the clash of symmetries has a physical Higgs field while the orbifold method
is Higgsless.} \cite{clash}. 

But many preparatory studies are needed
before such an ambitious vision can be realised. In this paper, we take
another important step along that road: the incorporation of gravity.
To facilitate the study of gravity, we begin by simplifying the group theoretic
structure as much as possible without losing the essence of the clash
of symmetries mechanism: we take $G=\ $U(1)$_1\otimes$U(1)$_2$, with
$H(-\infty)=\ $U(1)$_1$ and $H(+\infty)=\ $U(1)$_2$ leading to
$H_{\rm clash} = \{1\}$. The discrete symmetry we need for
topological stability interchanges the two Abelian sectors. 

In the next section we show that the ungauged version of this model can yield
a clash-of-symmetries-style kink in a warped metric spacetime.  The section after
that revisits the gauged extension in both flat and curved spacetime.

\section{The ungauged model}
\label{model_ungauged}

\subsection{The field equations}

The model has two complex Higgs fields $\Phi_1$ and $\Phi_2$ coupling to
$4+1$ dimensional gravity, and it respects the $\Phi_1 \leftrightarrow \Phi_2$
discrete symmetry. The action is
\begin{equation}
S = - \int \left[\frac{\kappa}{2}\, R + g^{MN}\, \left( \partial_M\overline{\Phi}_{1}\, \partial_N\Phi_{1}
+ \partial_M\overline{\Phi}_{2}\, \partial_N\Phi_{2} \right) 
+ V(\overline{\Phi}_{1}\Phi_1, \overline{\Phi}_{2}\Phi_2)\right]\, \sqrt{-g}\, d^5\, x
\end{equation}
where $g_{MN}$ is the $4+1$ dimensional metric tensor, with $M,N = 0,1,2,3,5$,
$R$ is the curvature scalar, and overbars denote complex conjugation. 
The matter action $S_m$ is defined as $S$
with the Einstein-Hilbert contribution absent.
The Higgs potential $V$ is as yet unspecified except that it must be invariant
under $\Phi_1 \leftrightarrow \Phi_2$. It will have a constant piece that serves
as the bulk cosmological constant. We employ the $(-,+,+,+,+)$ signature
and our sign conventions are as in Weinberg (apart from our definition of
$g \equiv {\rm Det}(g)$) \cite{weinberg}.

The Einstein equation found by requiring a stationary action under metric
variations is
\begin{equation}
G^{MN} = - \frac{1}{\kappa}\, T^{MN}
\end{equation}
where the Einstein tensor is, as usual, $G^{MN} = R^{MN} - (1/2) g^{MN} R$
with $R^{MN}$ being the Ricci tensor. The stress-energy tensor,
\begin{equation}
T^{MN} \equiv \frac{2}{\sqrt{-g}} \frac{\delta S_m}{\delta g_{MN}},
\end{equation}
evaluates to
\begin{equation}
T^{MN}= 2 g^{MP} g^{NQ} t_{PQ} - g^{MN} \left(g^{PQ} t_{PQ} + V \right),
\end{equation}
where
\begin{equation}
t_{MN} \equiv \partial_M\overline{\Phi}_{1}\, \partial_N\Phi_{1}
+ \partial_M\overline{\Phi}_{2}\, \partial_N\Phi_{2}.
\end{equation}
The Higgs Euler-Lagrange equations are
\begin{equation}
\partial_M\left( \sqrt{-g} g^{MN} \partial_N\Phi_{i} \right) =
\sqrt{-g}\, \frac{\partial V}{\partial \overline{\Phi}_{i}},
\end{equation}
where $i=1,2$, and the complex conjugate equations also hold.

We now specify our solution ansatz. We seek configurations
that only depend on the extra dimension $x^5 \equiv w$.
The imaginary parts of the Higgs fields
are set to zero and we write
\begin{equation}
\Phi_{i} = \frac{\phi_{i}(w)}{\sqrt{2}},
\end{equation}
where $\phi_i(w)$ is real. The metric ansatz, of Randall-Sundrum type, is defined through the 
$4+1$ dimensional line element
\begin{equation}
ds^2_5 = dw^2 + e^{2f(w)} ds^2_4
\end{equation}
where $f(w)$ is an as yet unknown function and $ds^2_4 =
\eta_{\mu \nu} dx^{\mu} dx^{\nu}$ is the $3+1$ dimensional
Minkowski line element (our brane will be Minkowski), with $\mu,\nu=0,1,2,3$ or $t,x,y,z$.
The field equations then reduce to
\begin{eqnarray}
3 f'' + 6 (f')^2 & = & - \frac{1}{\kappa} \left[ \frac{1}{2} \left[ ({\phi_{1}}')^2
+ ({\phi_{2}}')^2 \right] + V \right],\label{eq:00} \\
6 (f')^2 & = & \frac{1}{\kappa} \left[\frac{1}{2} \left[ ({\phi_{1}}')^2
+ ({\phi_{2}}')^2 \right] - V \right],\label{eq:ij}\\
{\phi_{i}}'' + 4 f' {\phi_{i}}' & = & \frac{\partial V}{\partial \phi_i},\label{eq:higgs}
\end{eqnarray}
where prime denotes differentiation with respect to $w$. Adding and subtracting
Eqs.\ (\ref{eq:00}) and (\ref{eq:ij}) gives
\begin{eqnarray}
f'' + 4(f')^2 & = & - \frac{2}{3\kappa} V,\label{eq:added}\\
f'' & = & - \frac{1}{3\kappa} \left[ ({\phi_{1}}')^2
+ ({\phi_{2}}')^2 \right].\label{eq:subtracted}
\end{eqnarray}
Note that Eq.\ (\ref{eq:subtracted}) is independent of the potential, a feature we will
use below.

We will construct a potential $V$ that has degenerate global minima at
\begin{equation}
{\rm vacuum\ 1:}\ \phi_1 = u,\ \phi_2 = 0\quad {\rm and}\quad 
{\rm vacuum\ 2:}\ \phi_1 = 0,\ \phi_2 = u,
\label{eq:vacua}
\end{equation}
where $u$ would be the VEV in the standard homogeneous SSB mechanism.
The vacua are related by the now spontaneously broken discrete interchange
symmetry. The Higgs field boundary conditions will be
\begin{equation}
\phi_i(-\infty) = {\rm vacuum\ 2}\quad {\rm and}\quad 
\phi_i(+\infty) = {\rm vacuum\ 1}.
\label{eq:bc}
\end{equation}

\subsection{Constructing the solution}

In $3+1$ dimensions without gravity, renormalisability limits us to
quartic potentials. With extra dimensions, and with gravity included, this is
no longer a well-motivated requirement. In fact, the only general constraint
on the form of $V$ we will impose is that it be bounded from
below. A very important topic for the future of the clash of symmetries
mechanism will indeed be the rules by which potentials should be
specified. 

Our more modest goal here is to provide an existence proof that kinks
featuring the clash of symmetries form can be consistently coupled to
Randall-Sundrum-like gravity. We will search for an analytically
tractable model for reasons of both simplicity and elegance. To do this, it makes
sense to invert the problem by first specifying the solutions
$\phi_i(w)$ and $f(w)$ and then constructing a potential that gives those
solutions. This, in itself, is non-trivial and interesting.

It turns out that a good choice, satisfying the boundary conditions of Eq.\ (\ref{eq:bc})
and also the Higgs-potential-independent Eq.\ (\ref{eq:subtracted}), is
\begin{eqnarray}
\phi_1^s(w) & = & \frac{u}{\sqrt{2}} \sqrt{1 + \tanh \beta w},\label{eq:phi1}\\
\phi_2^s(w) & = & \frac{u}{\sqrt{2}} \sqrt{1 - \tanh \beta w},\label{eq:phi2}\\
f^s(w) & = & - \frac{u^2}{12 \kappa} \ln \left( \cosh \beta w \right),\label{eq:f}
\end{eqnarray}
where $\beta$ is a free parameter that specifies the brane thickness, and the
superscript $s$ denotes ``solution''. 
The Higgs configurations are the square roots of the
most na\"{\i}ve choices, and $f^{s}(w)$ is essentially a regularisation of
$|w|$. The latter is, of course, the corresponding function in the 
Randall-Sundrum scenario. The smoothing out of the Randall-Sundrum cusp
goes hand-in-hand with the existence of Higgs field kinks coupled to gravity,
and the brane is dynamically generated rather than put in by hand.
Note also that
\begin{equation}
\phi_1^s(w)^2 + \phi_2^s(w)^2 = u^2,
\label{eq:circle}
\end{equation}
a feature that will turn out to be very useful.

The first inkling for what $V$ should be comes from Eq.\ (\ref{eq:added}).
It suggests that
\begin{equation}
V = - \frac{\beta^2 u^4}{24 \kappa} + \frac{\beta^2}{2 u^2}\, 
\left( 1 + \frac{u^2}{3\kappa} \right)\, \phi_1^2\, \phi_2^2 + 
U(\phi_1^2,\phi_2^2),
\label{eq:1stdraft}
\end{equation}
where $U$ is a function that vanishes at the level of the solution,
that is $U(\phi_1^s(w)^2,\phi_2^s(w)^2) = 0$. Notice
that a negative bulk cosmological constant,
\begin{equation}
\Lambda_5 = - \frac{\beta^2 u^4}{24 \kappa},
\end{equation}
follows from the ansatz: the bulk has anti-de Sitter qualities, as usual in Randall-Sundrum-like
models, though in our case it is of course not precisely anti-de Sitter.

More information about $U$ is supplied by the Higgs field equation (\ref{eq:higgs}).
To solve it, one must specify that
\begin{equation}
U = - \frac{\beta^2}{u^4}\, \left( \frac{3}{2} + \frac{u^2}{3\kappa} \right)\, 
\phi_1^2\, \phi_2^2\, (\, \phi_1^2 + \phi_2^2 - u^2\, ) + W(\phi_1^2 + \phi_2^2 - u^2),
\label{eq:2nddraft}
\end{equation}
where the new function $W$ obeys
\begin{eqnarray}
W(0) & = & 0,\label{eq:W1}\\
\left. \frac{\partial W}{\partial \phi_{1,2}} \right|_{\phi_1^s,\phi_2^s} & = & 0,
\label{eq:W2}
\end{eqnarray}
so as not to affect the satisfaction of the field equations, Eq.\ (\ref{eq:added})
and Eq.\ (\ref{eq:subtracted}) respectively. As far as they are concerned, $W$ can
vanish. The resulting sextic potential, however, would not be bounded from below
because the coefficient of the first term in $U$ is negative. 

We therefore construct
$W$ so that, as well as obeying Eqs.\ (\ref{eq:W1}) and (\ref{eq:W2}), it leads to
a potential that is both bounded from below and has global minima at Eq.\ (\ref{eq:vacua}).
The simplest suitable function is
\begin{equation}
W = \zeta\, \frac{\beta^2}{4u^2}\, \left( \frac{3}{2} + \frac{u^2}{3\kappa} \right)
(\, \phi_1^2 + \phi_2^2 - u^2\, )^2 \left( \eta + \frac{\phi_1^2 + \phi_2^2 - u^2}{u^2} \right)
\label{eq:3rddraft}
\end{equation}
where $\eta$ and $\zeta$ are dimensionless parameters.

The full Higgs potential for the model is specified by Eqs.\ (\ref{eq:1stdraft}),
(\ref{eq:2nddraft}) and (\ref{eq:3rddraft}). We will present a global minimisation analysis
of it in the next subsection.  (To reinstate the complex nature of the Higgs fields,
the substitution $\phi_i^2 \to 2 \overline{\Phi}_i \Phi_i$ should be made, though for the
forthcoming discussion we need only work with the $\phi_i$.)

Before doing so, let us identify the Randall-Sundrum limit of the model. We need to
identify a limit in which the warp factor, $e^{2f^s}$, becomes $e^{- a |w|}$ where $a$
is some positive constant. Following Ref.\cite{davidsonmannheim}, we note that $(\cosh \beta w)^{-(1/\beta)}
\to e^{-|w|}$ as $\beta \to \infty$. Writing
\begin{equation}
e^{2f^s(w)} = \left[ (\cosh \beta w)^{-\frac{1}{\beta}} \right]^{\frac{u^2\beta}{6\kappa}},
\end{equation}
we see that the Randall-Sundrum warp factor, $e^{- \frac{u^2\beta}{6\kappa}|w| }$, is obtained in the limit
\begin{equation}
\beta \to \infty,\quad u \to 0\quad {\rm such\ that}\ u^2 \beta \to {\rm finite\ constant}.
\label{eq:RSlimit}
\end{equation}
We also require $\eta$ and $\zeta$ to be unchanged when taking the limit.
Note that the Higgs field configurations, Eqs.\ (\ref{eq:phi1}) and (\ref{eq:phi2}), become
step functions as $\beta \to \infty$, with their magnitude tending to zero because $u \to 0$.
The Randall-Sundrum fine-tuning condition between the brane tension and the bulk cosmological constant
is a feature of our solution. Defining the brane tension via
\begin{equation}
\Lambda_b \equiv \lim_{RS} \int_{-\infty}^{+\infty}\, \left[ \frac{1}{2} \left[\, ({\phi^s_1}')^2
+ ({\phi^s_2}')^2\, \right] + V(\phi_1^s,\phi_2^s) \right]\, \sqrt{-g^s}\, dw,
\label{eq:Lambdabrane}
\end{equation}
it is easy to show that
\begin{equation}
\Lambda_b^2 = - \frac{3}{2} \kappa \Lambda_5,
\end{equation}
which is the Randall-Sundrum condition.

\subsection{Minimisation analysis}

We now identify the parameter space region where our potential has global minima
specified by Eq.\ (\ref{eq:vacua}). 

To establish the parameter region where the potential is bounded from below,
we look at the sixth-order terms in isolation. They are
\begin{equation}
V_6 = \lambda\, (\, \phi_1^2 + \phi_2^2\, ) \left[\,
\zeta\, (\, \phi_1^2 + \phi_2^2)^2 - 4\, \phi_1^2\, \phi_2^2\, \right],
\end{equation}
where $\lambda \equiv (3\beta^2/8u^4)[1 + 2u^2/(9\kappa)] > 0$. The 
substitutions $\phi_1 = \chi \cos\theta$ and $\phi_2 = \chi \sin\theta$
reveal that $V_6 = \lambda \chi^6 (\zeta - 4\sin^2\theta \cos^2\theta)$,
from which the condition
\begin{equation}
\zeta > 1
\label{eq:positivity}
\end{equation}
is trivially seen to be the only requirement to produce boundedness from below.

The minimisation analysis is a four-parameter problem. For some of
the local minima, both
their $(\phi_1,\phi_2)$ locations and the values of the potential
at those points are complicated functions of the parameters. But there
is a smart way to evade these unenlightening complications while
yielding physically meaningful results: approach the Randall-Sundrum limit
defined by Eq.\ (\ref{eq:RSlimit}). Our task then becomes a two parameter
problem in $(\eta, \zeta)$ that is readily visualised.

To see this, note that in Eqs.\ (\ref{eq:1stdraft}), (\ref{eq:2nddraft})
and (\ref{eq:3rddraft}), the combinations
\begin{equation}
1 + \frac{u^2}{3\kappa}\quad {\rm and}\quad \frac{3}{2} + \frac{u^2}{3\kappa}
\label{eq:combinations}
\end{equation}
frequently appear. In the Randall-Sundrum limit, which includes $u \to 0$, 
these combinations simplify to $1$ and $3/2$ respectively. In the exact 
Randall-Sundrum limit, the meaning to be ascribed to the potential is obscure,
so we cannot work exactly in that limit. Instead, we will approach the limit
without reaching it, and write an approximation to the potential justified
by the above remarks. It is
\begin{eqnarray}
V \simeq  V_{\rm approx} & = & 
- \frac{\beta^2 u^4}{24 \kappa} + \frac{\beta^2}{2 u^2}\, \phi_1^2\, \phi_2^2 
- \frac{3\beta^2}{2u^4}\, \phi_1^2\, \phi_2^2\, (\, \phi_1^2 + \phi_2^2 - u^2\, ) +
 \zeta\, \frac{3\beta^2}{8u^2}\, (\, \phi_1^2 + \phi_2^2 - u^2\, )^2 \left( \eta 
+ \frac{\phi_1^2 + \phi_2^2 - u^2}{u^2} \right),\nonumber\\
& = & \frac{3\beta^2u^2}{8}\, \left[ -\frac{u^2}{9\kappa} + \frac{4}{3}\, p_1^2\, p_2^2
- 4\, p_1^2\, p_2^2\, (\, p_1^2 + p_2^2 - 1\, )
+ \zeta\, (\, p_1^2 + p_2^2 - 1\, )^2 \left( \eta 
+ p_1^2 + p_2^2 - 1 \right) \right],
\label{eq:approxV}
\end{eqnarray}
where
\begin{equation}
p_i \equiv \frac{\phi_i}{u}
\end{equation}
is a dimensionless Higgs field. The fact that $\beta$ contributes only to an overall
scale, and that $\kappa$ appears solely in the constant term, reduces the analysis
to a two-parameter problem in $(\eta,\zeta)$. While $\kappa$ contributes to the
potential values, it does so equally for all local minima in this approximation,
which means that only $\eta$ and $\zeta$ need be considered when comparing the
local minima to see which are also the global minima.

The potential has three varieties, denoted $(I, II, III)$, of possible local minima (or saddle points),
\begin{eqnarray}
& (I)\quad & p_1^2  =  1,\ p_2=0\quad {\rm degenerate\ with}\quad p_1=0,\ p_2^2=1\quad ({\rm discrete\
symmetry\ breaking}),\label{eq:breaking}\\
& (II)\quad & p_1^2 = p_2^2 \neq 0\quad ({\rm discrete\ symmetry\ preserving}),\label{eq:preserving}\\
& (III)\quad & p_1 = p_2 = 0\quad ({\rm no\ SSB}).\label{eq:noSSB}
\end{eqnarray}
Which of these are actual local minima depends on $(\eta,\zeta)$.
We want $I$ to describe the global minima. 

\begin{figure}
\begin{center}
\epsfig{file=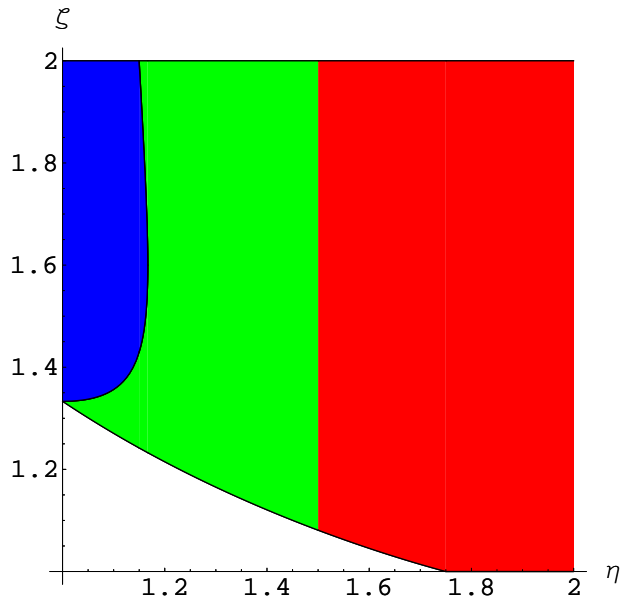,width=3.25in}
\caption{\label{fig:etazeta}
(Colour online.) The regions in the $(\eta,\zeta)$ plane where the approximate potential has the required
discrete symmetry breaking global minimum. The potential is bounded from below for $\zeta > 1$.
In the red (or medium-shaded) region, $\phi_1=\phi_2=0$
is a local maximum, $\phi_1^2 = \phi_2^2 \neq 0$ is a saddle point, and $\phi_1^2=u^2,\ \phi_2=0$
and $\phi_1=0,\ \phi_2^2=u^2$ are degenerate global minima. The green (or light-shaded) region differs
from the red (medium-shaded) only in that $\phi_1=\phi_2=0$ becomes a local minimum. As $\eta$ decreases
further, the $\phi_1=\phi_2=0$ minimum keeps falling and in the blue (or dark-shaded)
region it lies
below the $\phi_1^2 = \phi_2^2 \neq 0$ saddle point. For $\eta < 1$, $\phi_1 = \phi_2 = 0$
becomes the global minimum.
In the unshaded region in the $\eta,\zeta > 1$ quadrant, $\phi_1 = \phi_2 \neq 0$
is the global minimum. The border between the shaded and unshaded regions is a
smooth but complicated function $\zeta(\eta)$ given in the main text.
}
\end{center}
\end{figure}

Of the three, it
is $II$ that has the complicated parameter dependence.
In fact it is quite easy to show, even for the full potential of Eqs.\ (\ref{eq:1stdraft}),
(\ref{eq:2nddraft}) and (\ref{eq:3rddraft}), that $V(I) < V(III)$ requires that
\begin{equation}
\eta > 1
\end{equation}
which is thus a necessary but not sufficient condition to ensure that $I$ gives the
global minima. Sufficiency requires an examination of the parameter dependence of $V(II)$ also.
The results are shown in Fig.\ \ref{fig:etazeta}, where the shaded area is the Higgs parameter
region where $I$ is the global minimum.

In the red (or medium-shaded if viewing in black-and-white)
region, $I$ is a local minimum and $II$ a saddle point with $V(I) < V(II)$, while
$III$ is a local maximum. As $\eta$ decreases below $3/2$, $III$ develops into
a local minimum also; the ordering is $V(I) < V(II) < V(III)$ in the green (light-shaded) 
region
and $V(I) < V(III) < V(II)$ in the blue (dark-shaded). For $\eta < 1$, $III$ is the global minimum
so there is no SSB. The border line between the shaded and unshaded areas is
given by 
\begin{equation}
\zeta =
\frac{-27 - 18\,\eta + 61\,\eta^2 + {\sqrt{729 + 972\,\eta - 2970\,\eta^2 + 1900\,\eta^3 - 375\,\eta^4}}}{24\,\eta^3}.
\end{equation}
Lowering $\zeta$ has the effect of lowering $V(II)$. Below this border line,
$V(II) < V(I) < V(III)$, so the symmetry preserving vacuum is the global minimum.

The border between the blue (dark-shaded) and green (light-shaded) regions is given by
\begin{equation}
\eta = 
\frac{8\,\zeta - 3\,\zeta^2 + 2\,{\sqrt{3}}\,{\sqrt{4\,\zeta^2 - 7\,\zeta^3 + 3\,\zeta^4}}}{3\,\zeta^2}.
\end{equation}

\begin{figure}
\begin{center}
\epsfig{file=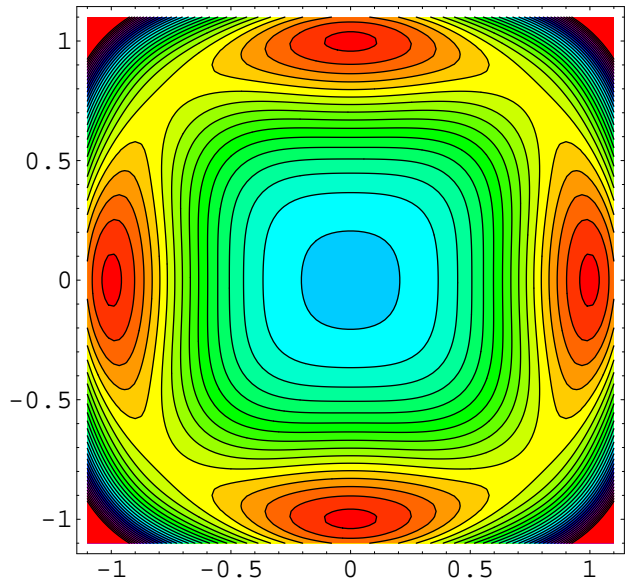,width=3.25in}
\caption{\label{fig:red}
(Colour online.) Contour plot of the approximate potential for $\eta=\zeta=1.6$ to illustrate
the qualitative features of the extrema for a representative point in
the red (medium-shaded) region of Fig.\ \ref{fig:etazeta}. Within this plot, red colour codes
for minima and blue for maxima. (These look similar in black-and-white, but the description
below clarifies which is which.)
The origin, $p_1 = p_2 = 0$, is a local maximum.
The degenerate global minima are at $(p_1,p_2)=(0,\pm 1)$ and $(\pm 1,0)$,
while extrema $II$, where $p_1^2=p_2^2$, are saddle points (light green or very-lightly shaded). 
}
\end{center}
\end{figure}

\begin{figure}
\begin{center}
\epsfig{file=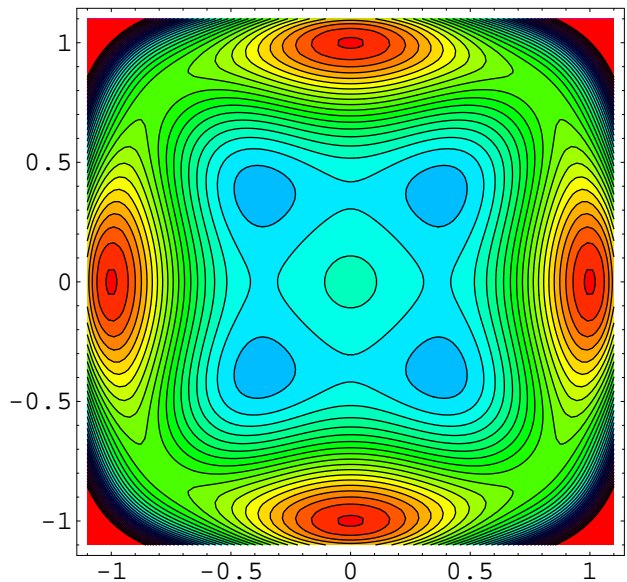,width=3.25in}
\caption{\label{fig:green}
(Colour online.) Contour plot of the approximate potential for $\eta=1.3$ and $\zeta=1.6$ to illustrate
the qualitative features of the extrema for a representative point in
the green (light-shaded) region of Fig.\ \ref{fig:etazeta}. Within this plot, red colour codes
for minima and blue for maxima. (In black-and-white, the red region appears darker than the blue
in this plot; the description below clarifies which is which.)
The origin, $p_1 = p_2 = 0$, is now a local minimum,
but the degenerate global minima are still at $(p_1,p_2)=(0,\pm 1)$ and $(\pm 1,0)$.
The relevant $p_1^2=p_2^2$ extrema are again saddle points. 
}
\end{center}
\end{figure}

\begin{figure}
\begin{center}
\epsfig{file=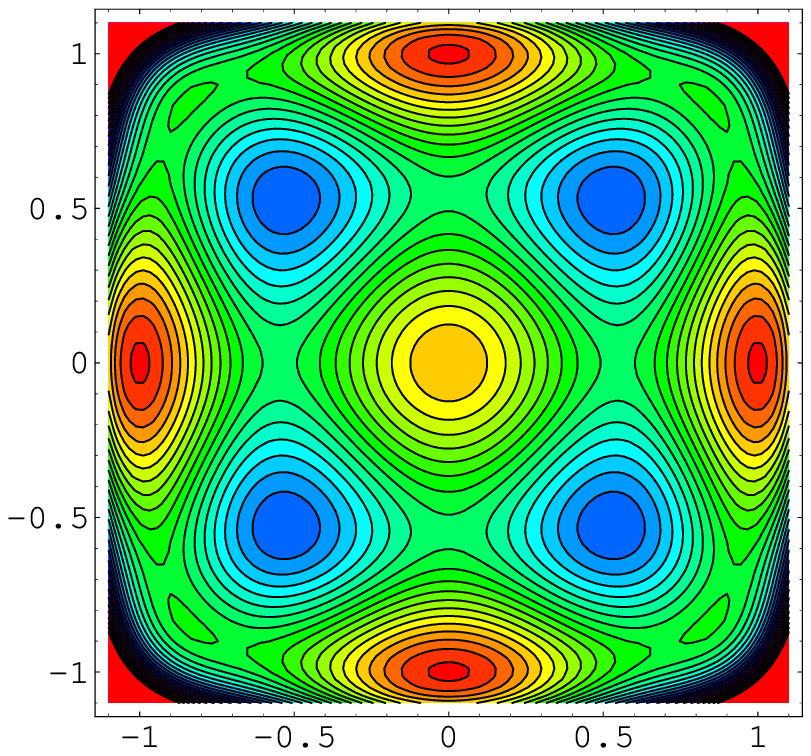,width=3.25in}
\caption{\label{fig:blue}
(Colour online.) Contour plot of the approximate potential for $\eta=1.05$ and $\zeta=1.6$ to illustrate
the qualitative features of the extrema for a representative point in
the blue (dark-shaded) region of Fig.\ \ref{fig:etazeta}. Within this plot, red colour codes
for minima and blue for maxima. (In black-and-white they appear similar, but the
description below clarifies which is which.)
The origin, $p_1 = p_2 = 0$, is now a deeper local 
minimum (yellow, or medium-shaded), having dropped below the $p_1^2=p_2^2$ saddle points (green, or
medium-shaded valleys). The local maxima are at approximately $(|p_1|, |p_2|) = (0.5, 0.5)$.
The degenerate global minima are at $(p_1,p_2)=(0,\pm 1)$ and $(\pm 1,0)$.
}
\end{center}
\end{figure}

Figures \ref{fig:red}-\ref{fig:blue} show contour plots of the approximate potential
for representative points in the red (medium-shaded), green (light-shaded) and 
blue (dark-shaded) areas of Fig.\ \ref{fig:etazeta}, respectively.
The colour- or shade-coding is explained in the captions. The plots illustrate the pattern
of extrema described above.

\section{The gauged model}
\label{model_gauged}

It is natural to extend our toy model by gauging U(1)$\otimes$U(1), and to study
the dependence of the gauge fields on $w$ by self-consistently solving all of the
coupled Euler-Lagrange equations.  In the flat space case, Rozowsky, Volkas and Wali \cite{gaugeclash}
found some interesting solutions where the gauge field coupled to $\Phi_i$ 
tended towards a linear function of $w$ on the side of the wall where $\Phi_i \to 0$,
and was exponentially suppressed on the other side (Meissner effect).  The
linearly rising gauge field implies an asymptotically constant magnetic field,
so the physical picture was of two infinite, uniform sheets (one for each sector)
of supercurrent density flowing along the domain wall.  In this section we will
first generalise the Rozowsky et al.\ flat-space analysis and then consider
the curved space problem.

\subsection{Flat space}

The Lagrangian is given by
\begin{equation}
{\cal L} = - \frac{1}{4} F_{1}^{MN} F_{1MN} - \frac{1}{4} F_{2}^{MN} F_{2MN} 
- ( D^{M} \Phi_1 )^{*}( D_{M} \Phi_1 ) - ( D^{M} \Phi_2 )^{*}( D_{M} \Phi_2 ) - V,
\end{equation}
where
\begin{eqnarray}
D_{M} \Phi_1 & = & ( \partial_{M} - i e A_{1M} - i \tilde{e} A_{2M} )\Phi_1,\nonumber \\
D_{M} \Phi_2 & = & ( \partial_{M} - i \tilde{e} A_{1M} - i e A_{2M} ) \Phi_2,
\end{eqnarray}
and we take $V$ to be quartic
\begin{equation}
V = \lambda_1 (\overline{\Phi}_1 \Phi_1 + \overline{\Phi}_2 \Phi_2 - u^2)^2 
+ \lambda_2 \overline{\Phi}_1 \Phi_1 \overline{\Phi}_2 \Phi_2.
\end{equation}
In the $\lambda_{1,2} > 0$ parameter space region, the global minima of $V$ are
\begin{equation}
{\rm Vacuum\ 1}:\ \overline{\Phi}_1 \Phi_1 = u^2,\ \Phi_2 = 0\quad
{\rm and}\quad {\rm Vacuum\ 2}:\ \Phi_1 = 0,\  \overline{\Phi}_2 \Phi_2 = u^2.
\end{equation} 
The Lagrangian is invariant under the discrete symmetry
\begin{equation}
\Phi_1 \leftrightarrow \Phi_2,\qquad A_1 \leftrightarrow A_2.
\end{equation}
This model generalises that considered in Ref.\cite{gaugeclash} by having $\tilde{e} \neq 0$.

The Euler-Lagrange equations are
\begin{eqnarray}
D^M D_M \Phi_i & = & 2\, \lambda_1\, \Phi_i\, (\, \overline{\Phi}_i \Phi_i + \overline{\Phi}_j \Phi_j - u^2\, )
+ \lambda_2\, \Phi_i\, \overline{\Phi}_j\, \Phi_j,\nonumber \\
\partial_M\, F_{i}^{MN} & = & - 2\, {\rm Im}\,(\, e \overline{\Phi}_i\, D^N\, \Phi_i 
+ \tilde{e}\, \overline{\Phi}_j\, D^N\, \Phi_j\, ),
\end{eqnarray}
where $(i,j) = (1,2)$ or $(2,1)$.  We now specialise to configurations that
depend only on $w$, impose Lorentz gauge $\partial_M\, A_{i}^{M} = 0$, and write
$\Phi_i = R_i\, e^{i\theta_i}$.  The equations become,
\begin{eqnarray}
R''_i & = & R_i\, \left[\, (e\, A_{i}^{\mu} + \tilde{e}\, A_{j}^{\mu}\, )\, 
(e\, A_{i\mu} + \tilde{e}\, A_{j\mu}\, )\, \right] 
+ 2\, \lambda_1\, R_i\, (\, R_i^2 + R_j^2 - u^2\, ) + \lambda_2\, R_i\, R_j^2,\label{eq:R}\\
{A_{i}^{\mu}}'' & = & 2\, e\, R_i^2\, (\, e\, A_{i}^{\mu} + \tilde{e}\, A_{j}^{\mu}\, ) 
+ 2\, \tilde{e}\, R_j^2\, (\, e\, A_{j}^{\mu} + \tilde{e}\, A_{i}^{\mu}\, ),\label{eq:gauge}\\
\theta'_i & = & - (\, e\, A_{i}^{5} + \tilde{e}\, A_{j}^{5}\, ). \label{eq:puregauge}
\end{eqnarray}
Observe from the last of these equations that the $A=5$ gauge field components are 
pure gauge, so they and the phase fields $\theta_i$ decouple from the problem.

We choose to look for solutions that 
do not break rotational invariance in the $x$, $y$
and $z$ directions.  This means that we can always rotate the $x,y,z$ coordinate
system so that the only nonzero component of $(A_{i}^{x}, A_{i}^{y}, A_{i}^{z})$ is in, say,
the $x$-direction (this is tantamount to the requirement that the ratios of these
components are $w$-independent).  We also choose that the same component is nonzero
for both $i=1$ and $i=2$.  Equations (\ref{eq:R}) and (\ref{eq:gauge}) now simplify to
\begin{eqnarray}
R''_i & = & R_i\, \left[\, - (e\, A_{i}^{t} + \tilde{e}\, A_{j}^{t})^2 
+ (e\, A_{i}^{x} + \tilde{e}\, A_{j}^{x})^2\, \, \right] 
+ 2\, \lambda_1\, R_i\, (\, R_i^2 + R_j^2 - u^2\, ) + \lambda_2\, R_i\, R_j^2,\label{eq:modR}\\
{A_{i}^{t,x}}'' & = & 2\, e\, R_i^2 (\, e\, A_{i}^{t,x} + \tilde{e}\, A_{j}^{t,x}\, )
+ 2\, \tilde{e}\, R_j^2 (\, e\, A_{j}^{t,x} + \tilde{e}\, A_{i}^{t,x}\, )\label{eq:modgauge}
\end{eqnarray}
The final choice is whether to choose the gauge fields to be timelike $(A_{i}^{t}\neq 0,A_{i}^{x}=0)$, 
lightlike $(A_{i}^{t} = A_{i}^{x} \neq 0)$, or spacelike $(A_{i}^{t} = 0,A_{i}^{x} \neq 0)$.  
As is obvious from Eq.\ (\ref{eq:modR}), the timelike ansatz gives rise to asymptotic oscillatory
behaviour for the $R$'s and is thus unacceptable.  We first discuss the spacelike case.

\begin{figure}
\centering
\includegraphics[width=0.49\textwidth]{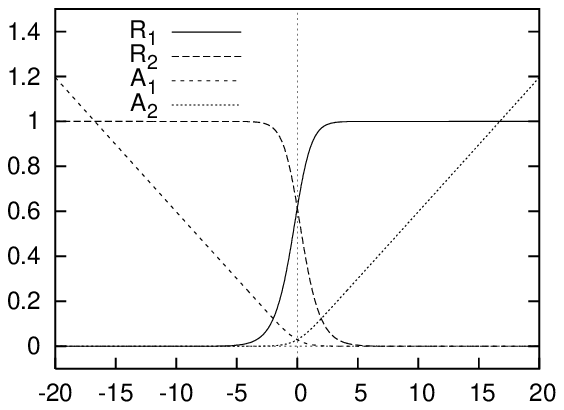}
\hfill
\includegraphics[width=0.49\textwidth]{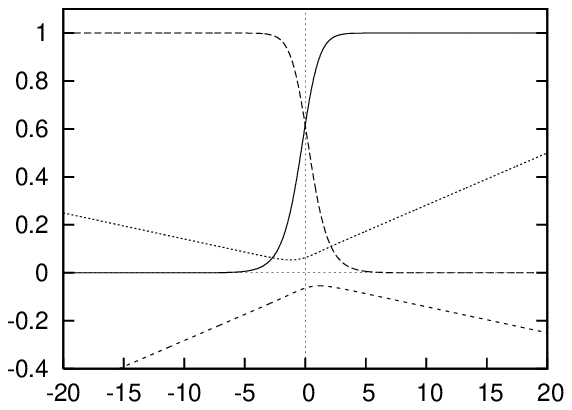}
\includegraphics[width=0.49\textwidth]{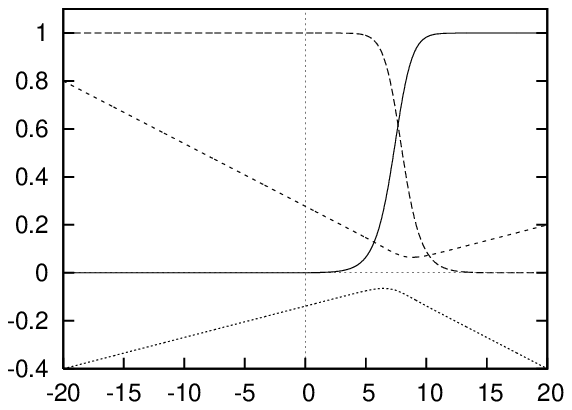}
\hfill
\includegraphics[width=0.49\textwidth]{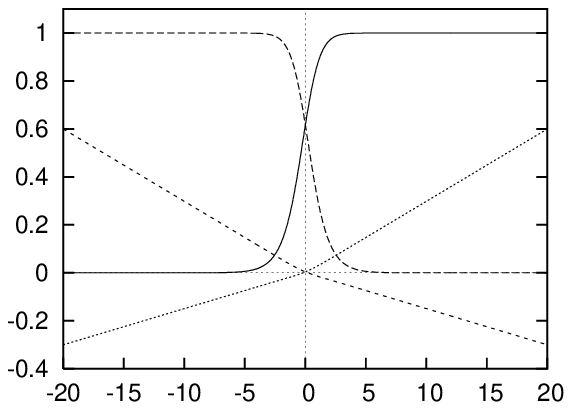}
\caption{
\small
Static solutions to the two scalar and two gauge fields in the
U(1)$\otimes$U(1) model.  In all cases $\lambda_1 = \lambda_2 = e = 1$,
and the fields are all plotted in units of $u$.
In the top left plot, $\tilde{e}=0$ as in the Rozowsky et al.\
model and the gauge fields show Meissner-like suppression under
their respective scalar field.  In the other three plots,
$\tilde{e}=\tfrac{1}{2}$ and we show cases with various values
for the free gauge field boundary conditions.  In the two right
plots, reflection symmetric boundary conditions are imposed for the
pair of linear combinations $e A_i^x + \tilde{e} A_j^x$.  In the
bottom left plot, the gauge boundary conditions are asymmetric and
their coupling to the kink shifts it to the right.
}
\label{fig:higgs-gauge-static}
\end{figure}

As a boundary condition, we require that $R_{1,2}$ tend to vacua 1 and 2 on opposite sides
of the wall:
\begin{equation}
R_1(+\infty) = u,\ R_2(+\infty) = 0\quad {\rm and}\quad
R_1(-\infty) = 0,\ R_2(-\infty) = u.
\end{equation}
The appropriate boundary conditions for the gauge fields are then determined by examining
Eqs.\ (\ref{eq:modgauge}) at the asymptotic points.  As $w \to +\infty$, we find that 
\begin{equation}
\left( \begin{array}{c} {A_1}'' \\ {A_2}'' \end{array} \right)
= 2\, u^2\, \left( \begin{array}{cc}
e^2 & e\, \tilde{e} \\ e\, \tilde{e} & \tilde{e}^2 \end{array} \right)
\left( \begin{array}{c} A_1 \\ A_2 \end{array} \right),
\end{equation}
where $A_i \equiv A_i^x$.
Inputting trial asymptotic solutions of the form
\begin{equation}
A_{1,2} = a_{1,2}\, e^{-\kappa w},
\end{equation}
we obtain an eigenvalue equation yielding
\begin{equation}
\kappa = \sqrt{2}\, u\, \sqrt{e^2 + \tilde{e}^2}\quad {\rm and}\quad \kappa = 0.
\end{equation}
The zero eigenvalue means that one linear combination of $A_1$ and $A_2$ is always 
unsuppressed, 
and the vanishing second derivative tells us it grows linearly with $w$ (provided its
amplitude is nonzero).  
By the symmetry of the problem, a similar situation obtains as $w \to -\infty$.
The exponentially suppressed linear combination embodies the Meissner effect. Numerical
solutions are depicted in Fig.\ \ref{fig:higgs-gauge-static}.

The lightlike case is qualitatively similar to the spacelike case.  But
it is amusing to point out that for the parameter choice $\lambda_2=4\lambda_1$
an analytic solution exists. For the lightlike ansatz, the gauge fields
cancel out of Eq.\ \ref{eq:modR}, and with the stated parameter choice
and $e=1$, $\tilde{e}=0$, the resulting equations yield the solutions
\begin{equation}
R_{1,2}(w) = \frac{u}{2}\,\left[1 \pm \tanh(\sqrt{\lambda_1}\, u\, w)\right].
\end{equation}
Inputting these results into Eq.\ \ref{eq:modgauge}, the analytic solutions
obeying the correct boundary conditions
for the gauge field functions $A_{i}^{t} = A_{i}^{x} \equiv A_i$ are
\begin{equation}
A_{1,2}(w) = {\rm const.}\times {_2F_1}\left(\, \tfrac{1+\sqrt{2}-\sqrt{3}}{2}\, ,\, \tfrac{1+\sqrt{2}+\sqrt{3}}{2}\, ;\, 
1 + \sqrt{2}\, ;\, (1 + e^{\pm2\sqrt{\lambda_1}uw})^{-1}\, \right)\, 
\left(1 \mp \tanh(\sqrt{\lambda_1}\, u\, w)\right)^{\frac{1}{\sqrt{2}}},
\end{equation}
where $_2F_1$ is a hypergeometric function.

\subsection{Warped spacetime}

We now generalise the above model through the inclusion of gravity \cite{gaugeggrav}. The action is
\begin{equation}
S = - \int \left [ \frac{\kappa}{2} R
                   + g^{M N} t_{M N}
                   + \tfrac{1}{4} g^{M P} g^{N Q} f_{M N P Q}
                   + V
           \right ] \sqrt{-g} \; \md^5 x,
\end{equation}
with
\begin{eqnarray}
t_{M N}     &\equiv& \sum_i \left ( D_M \Phi_i \right ) ^* D_N \Phi_i, \\
D_M         &\equiv& \partial_M - \sum_i \mi Q_i A_{i M}, \\
f_{M N P Q} &\equiv& \sum_i F_{i M N} F_{i P Q},
\end{eqnarray}
and the charges $(Q_1,Q_2)$ of the scalar fields are
$(e, \tilde{e})$ for $\Phi_1$ and $(\tilde{e}, e)$ for $\Phi_2$.

The Higgs and gauge field equations of motion are
\begin{eqnarray}
D_M \left ( \sqrt{-g} \; g^{M N} D_N \Phi_i \right ) - \sqrt{-g} \frac{\partial V}{\partial \Phi_i^*} &=& 0 \\
\partial_M \left ( \sqrt{-g} \; g^{M P} g^{N Q} F_{i P Q} \right )
+ \sqrt{-g} \; g^{N P} 2 \left ( e \imag \left ( \Phi_i^* D_P \Phi_i \right )
                              + \tilde{e} \imag \left ( \Phi_j^* D_P \Phi_j \right ) \right ) & =& 0,
\end{eqnarray}
while the Einstein equations, $\kappa\, G_{M N} = - T_{M N}$, feature
\begin{eqnarray}
T_{M N} &=& 2 t_{M N} + g^{P Q} f_{M P N Q} + g_{M N} \mcl_M, \\
\mcl_M  &=& -g^{M N} t_{M N} - \tfrac{1}{4} g^{M P} g^{N Q} f_{M N P Q} - V.
\end{eqnarray}

The metric ansatz is generalised to
\begin{equation}
\md s^2 = - \me^{f(w)} \md t^2
       + \me^{h(w)} \md x^2
       + \me^{j(w)} \left ( \md y^2 + \md z^2 \right )
       + \md w^2,
\end{equation}
in order to have enough degrees of freedom to be able to look at sufficiently general
gauge field configurations.  We look for solutions of the form
\begin{eqnarray}
A_{i M} &=& \left ( A_i(w), B_i(w), 0, 0, Z_i(w) \right ), \\
\Phi_i  &=& R_i(w) \me^{\mi \alpha_i(w)},
\end{eqnarray}
To simplify the algebra it is useful to define the following:
\begin{eqnarray}
\mcf                       &=& \tfrac{1}{2} \left ( f + h + 2 j \right ), \\
\overline{\mcf}            &=& \tfrac{1}{2} \left ( -f + h + 2 j \right ), \\
\overline{\overline{\mcf}} &=& \tfrac{1}{2} \left ( f - h + 2 j \right ), \\
\mca_i &=& e A_i + \tilde{e} A_j, \\
\mcb_i &=& e B_i + \tilde{e} B_j, \\
\mcz_i &=& e Z_i + \tilde{e} Z_j - \alpha_i' .
\end{eqnarray}
In terms of these, the equations of motion for the gauge field components are
\begin{eqnarray}
A_i''
  + \overline{\mcf}' A_i'
  - 2 e R_i^2 \mca_i
  - 2 \tilde{e} R_j^2 \mca_j &=& 0, \label{eq:wA}\\
B_i''
  + \overline{\overline{\mcf}}' B_i'
  - 2 e R_i^2 \mcb_i
  - 2 \tilde{e} R_j^2 \mcb_j &=& 0, \label{eq:wB}\\
\left ( e R_i^2 + \tilde{e} R_j^2 \right ) \mcz_i &=& 0. \label{eq:wZ}
\end{eqnarray}
Equation (\ref{eq:wZ}) implies $\mcz_i=0$ which we use to simplify
the rest of the equations of motion.  For the scalar fields we
have
\begin{equation}
R_i''
  + \mcf' R_i'
  + \me^{-f} \mca_i^2 R_i
  - \me^{-h} \mcb_i^2 R_i
  - \me^{-\mi \alpha_i} \frac{\partial V}{\partial \Phi_i^*} = 0,
\label{eq:wR}
\end{equation}
and for the three metric functions we have
\begin{eqnarray}
f''
  + \tfrac{1}{2} f'^2
  + \tfrac{1}{3} f' h'
  + \tfrac{2}{3} f' j'
  - \tfrac{1}{3} h' j'
  - \tfrac{1}{6} j'^2
  - \tfrac{10}{3 \kappa} \me^{-f} \Phi
  - \tfrac{2}{3 \kappa} \me^{-h} \Psi
  + \tfrac{2}{3 \kappa} \Omega &=& 0, \label{eq:wf} \\
h''
  + \tfrac{1}{3} f' h'
  - \tfrac{1}{3} f' j'
  + \tfrac{1}{2} h'^2
  + \tfrac{2}{3} h' j'
  - \tfrac{1}{6} j'^2
  + \tfrac{2}{3 \kappa} \me^{-f} \Phi
  + \tfrac{10}{3 \kappa} \me^{-h} \Psi
  + \tfrac{2}{3 \kappa} \Omega &=& 0, \label{eq:wh} \\
j''
  - \tfrac{1}{6} f' h'
  + \tfrac{1}{6} f' j'
  + \tfrac{1}{6} h' j'
  + \tfrac{5}{6} j'^2
  + \tfrac{2}{3 \kappa} \me^{-f} \Phi
  - \tfrac{2}{3 \kappa} \me^{-h} \Psi
  + \tfrac{2}{3 \kappa} \Omega &=& 0, \label{eq:wj}
\end{eqnarray}
where
\begin{eqnarray}
\Phi &=& \mca_1^2 R_1^2 + \mca_2^2 R_2^2
          + \tfrac{1}{2} A_1'^2 + \tfrac{1}{2} A_2'^2, \\
\Psi &=& \mcb_1^2 R_1^2 + \mcb_2^2 R_2^2
          + \tfrac{1}{2} B_1'^2 + \tfrac{1}{2} B_2'^2, \\
\Omega &=& R_1'^2 + R_2'^2 + V.
\end{eqnarray}

We performed an extensive numerical search for solutions to 
Eqs.\ (\ref{eq:wA},\ \ref{eq:wB},\ \ref{eq:wR}-\ref{eq:wj}), subject to the
boundary conditions that the Higgs fields tend to VEVs, and that the
metric functions tend to Randall-Sundrum form.  For definiteness, we used
the same Higgs potential as in the non-gauged model, namely as given by
Eqs.\ (\ref{eq:1stdraft}), (\ref{eq:2nddraft}) and (\ref{eq:3rddraft}).
The result of this investigation was the perhaps 
disappointing one that all solutions found
required the gauge fields to vanish.  Since the investigation was numerical,
we cannot be certain that no non-trivial solutions exist, although we
strongly suspect this to be the case.

\section{Conclusion}
\label{conclusion}

We have constructed a U(1)$\otimes$U(1) 
model with two complex scalar fields interchanged by a discrete symmetry 
coupled to $4+1$
dimensional gravity. It has a solution featuring 
clash-of-symmetries-style Higgs
kink configurations in a Randall-Sundrum-like spacetime. Gravity is
localised to the dynamically generated (smooth) brane, and
the Randall-Sundrum limit of the solution can be defined. 
For the chosen Higgs kink and metric configurations, the Higgs
potential had to be of a certain sextic form whose properties we
studied in some depth.
The symmetry breaking pattern varies as a function of the extra dimension
coordinate $w$, and displays the clash of symmetries phenomenon:
at all points $|w|<\infty$ both U(1)'s are broken, with alternate
U(1)'s restored as $w \to +\infty$ and $w \to -\infty$.
The spontaneous breaking of the discrete interchange symmetry
guarantees topological stability for the Higgs-gravity-induced brane.
We also examined the gauged version of the theory, but we found that
to have nonzero gauge fields the spacetime has to be flat.

This work sets the stage for incorporating gravity into more
complicated models displaying the clash of symmetries. The
eventual aim is to construct a realistic brane-world model that
uses the clash of symmetries to understand spontaneous symmetry
breaking in a more satisfactory fashion than obtained with
the default mechanism using homogeneous Higgs vacuum expectation values.

\acknowledgments{GD was supported in part by the Commonwealth of Australia,
DPG by the Puzey bequest to the University of Melbourne,
RRV by the Australian Research Council, and KCW in part by the U.S.\
Department of Energy (DOE) grant DE-FG02-85ER40237. 
RRV would like to thank D. Vignaud,
A. Goldwurm and the APC group based at the Coll\`{e}ge de France for their
hospitality while this manuscript was completed.}

\end{document}